# Tuning light emission crossovers in atomic-scale aluminum plasmonic tunnel junctions

*Yunxuan Zhu[1], Longji Cui[2,3], Mahdiyeh Abbasi[4] and Douglas Natelson[1,4,5,*]*

[1]Department of Physics and Astronomy, Rice University, Houston, TX 77005, United States

[2]Paul M. Rady Department of Mechanical Engineering, University of Colorado, Boulder, CO 80309, United States.

[3]Materials Science and Engineering Program, University of Colorado, Boulder, CO 80309, United States.

[4]Department of Electrical and Computer Engineering, Rice University, Houston, TX 77005, United States.

[5]Department of Materials Science and Nanoengineering, Rice University, Houston, TX 77005, United States.

KEYWORDS: Plasmonics, nanoscale junction, atomic sized light emission, hot-carrier dynamics, electron inelastic tunneling

**ABSTRACT**

Atomic sized plasmonic tunnel junctions are of fundamental interest, with great promise as the smallest on-chip light sources in various optoelectronic applications. Several mechanisms of light emission in electrically driven plasmonic tunnel junctions have been proposed, from single-electron or higher order multi-electron inelastic tunneling to recombination from a steady-state population of hot carriers. By progressively altering the tunneling conductance of an aluminum junction, we tune the dominant light emission mechanism through these possibilities for the first time, finding quantitative agreement with theory in each regime. Improved plasmonic resonances in the energy range of interest increase photon yields by two orders of magnitude. These results demonstrate that the dominant emission mechanism is set by a combination of tunneling rate, hot carrier relaxation timescales, and junction plasmonic properties.

Plasmon enhanced atomic scale light sources are of great interest with potential optoelectronic applications to bridge the bandwidth mismatch between electronics and photonics. An electrically biased light-emitting tunnel junction provides a feasible platform to investigate light matter interaction at the atomic scale, particularly the role of plasmonic excitations with deep subwavelength confinement. In the low current limit, electrons may inelastically tunnel through the barrier, exciting a localized surface plasmon (LSP) individually which then with some probability radiatively decays to emit a photon, with the upper photon energy threshold set by the applied voltage bias ($\hbar\omega \leq eV_b$) [1–7]. In the high current limit, however, photon emission above the voltage threshold ($\hbar\omega > eV_b$) has been observed. The underlying physics of the above-threshold photon emission remains an open question. Besides thermal broadening due to the finite temperature of the electrically driven tunnel junction, the excess photon energy can also be achieved either through higher order multielectron coherent interactions [8–16], or plasmon-generated[17–19] or Joule heating-produced[20–22] hot carriers. The above-mentioned candidate mechanisms are all proposed in different experimental platforms, which makes it hard to distinguish each contribution to the total above-threshold photons. To demystify the underlying mechanism and potentially resolve the current debate in this field, it is desirable to study all the possible mechanisms in the same system under different parameter regimes.

Previous works based on either silver or gold tunnel junctions in scanning tunneling microscopes (STM) or using planar nanostructures have demonstrated either multielectron-based or hot carrier-induced above-threshold emission[13,14,19,21,23]. Compared to silver and gold, aluminum stands out as a strong, malleable plasmonic material with a larger damping of plasmons and lack of interband transition within the visible spectrum energy range. These characteristics can potentially suppress the hot carrier generation channel and eliminate complicated carrier transitions through different bands during the inelastic electron tunneling process, making Al a good candidate for probing above-threshold light emission while possessing excellent malleability and distinct electronic structure[24–26]. Furthermore, due to aluminum's malleability, compared to gold it is easier to further electromigrate the Al into the conductance range favoring the single and multielectron tunneling transport.

In this paper, leveraging multi-step electromigration of Al tunnel junctions, we show that the effective interelectrode gap size and the tunneling current magnitude can be systematically varied, which enables us to tune the light emission mechanisms through experimental parameter domains to observe the crossover between below- and above-threshold light emission. Furthermore, we realize the first-ever experimental demonstration of the crossover between the multi-electron transport to plasmonic hot carrier-induced above-threshold photon emission. Our observations of light emission in different current regimes agree quantitatively well with theoretical models. These results demonstrate that the dominant emission mechanism is set by a combination of tunneling rate, hot carrier relaxation timescales, and junction plasmonic properties, which could potentially resolve the long-standing debate regarding the relative roles of these candidate mechanisms. Electrically controlled plasmonic light emission platforms set the stage for tunable atomic scale plasmonic light sources that leverage different electron and plasmonic dynamics in various practical applications.

Fig. 1a, b and c summarize the mechanisms contributing to both above-threshold and below-threshold light emission from a biased tunnel junction. In Fig. 1a, electrons can inelastically tunnel through the barrier, exciting a localized surface plasmon (LSP), which subsequently decay radiatively to produce photon emission. For a purely first order process (i.e. each tunneling electron may be treated individually), photon energy is limited by the applied bias. At higher currents, higher order processes are also possible, in which two or more electrons can tunnel coherently (Fig. 1b), resulting in above-threshold emission ($\hbar\omega > eV_b$), though with a significantly lower probability. As a consequence, the quantized nature of those events is embedded in the spectra and can be discerned if the wavelength dependent plasmon enhancement is removed, an approach commonly employed to verify this mechanism[13,27]. Note that this analysis approach assumes featureless electronic densities of states for the source/drain electrodes over the bias window, an assumption that is reasonable at biases below ~ 1.5 V[28]; nontrivial electronic structure of the junction electrodes would lead to additional energy dependences[7].

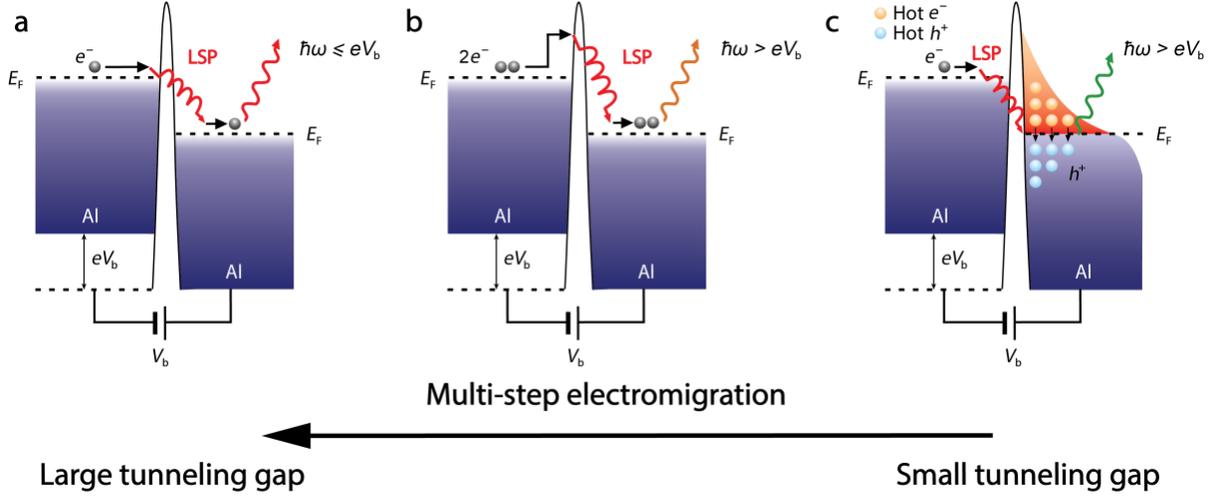

**Figure 1.** Schematics for various light emission mechanisms and the electromigration experimental setup. (a) an electron can inelastically tunnel from source to the drain, exciting a localized surface plasmon (LSP) which subsequently decays into a radiative photon with energy limited by the LSP ($\hbar\omega \leq eV_b$) (b) Two electrons can tunnel through the junction barrier coherently at a lower probability, creating a higher energy LSP and decaying in the same way to generate above-threshold emission ($\hbar\omega > eV_b$). (c) Instead of decaying through radiative emission, the electrically excited LSP decays non-radiatively, exciting hot electrons and hot holes above and below the Fermi energy, which recombine radiatively to produce above-threshold photons. Note that the hot electrons and holes are generated in *both* the source and drain electrodes due to the plasmonic decay process. The schematic showing this in only one side is just for clarity.

Rather than directly decaying to photons, these electrically excited plasmons can instead go through non-radiative decay channels and, at sufficiently high currents, form a steady-state population of nonequilibrium hot carriers far away from the Fermi level in both source and drain electrodes, which subsequently recombine on a tens of fs time scale[29,30] and radiate above-threshold photons (Fig. 1c). To sustain a steady-state hot carrier contribution to the electronic distribution through this mechanism, high currents (typically over tens of μA) and strong LSP resonances are necessary. Electromigrated planar junctions can support an intense transverse dipolar plasmon that hybridizes with the highly localized LSP modes in the nanogap[31], which favors the condition for hot carrier generation.

For clarity, we will first present the light emission results for a specific Al based junction from below-threshold (single electron inelastic tunneling) to above-threshold (multielectron cotunneling or hot carrier recombination), which is the reverse order of the electromigration process. Statistical analysis for the junction conductance after each step electromigration is shown in section 1 of the Supporting Information (SI). Because it has proven straightforward to migrate finely through different conductance levels, Al is a more suitable material compared to Au to tune the conductance in the single and multi-electron inelastic tunneling regime. The below-threshold light emission for an Al junction with largest gap distance after a third step of electromigration ($1.5 \times 10^{-3} G_0$ zero bias conductance, where $G_0 \equiv 2e^2/h$) is shown in Fig. 2a, where all overbias emission is suppressed with a sharp decrease at the bias threshold cutoff. Such emission can be well explained by the single electron inelastic tunneling model. To illustrate this mechanism, we numerically calculate the emission spectrum by inserting the zero bias junction conductance and $\rho(\hbar\omega)$ (extracted from light emission at an earlier stage of electromigration) into the formula. The total spectrum as a function of bias and photon energy $U_{total}(V_b, \hbar\omega)$ can be expressed by adding all contributions from different orders of tunneling processes, $U_{total}(V_b, \hbar\omega) = U_{1e}(V_b, \hbar\omega) + U_{2e}(V_b, \hbar\omega) \cdots$ (See section 3 of the SI for calculation details.). The $1e$, $2e$ labels indicate different order tunneling processes. Due to the low conductance here, contributions from higher order processes are negligible compared to $U_{1e}(V_b, \hbar\omega)$,

$$U_{1e}(V_b, \hbar\omega) = \rho(\hbar\omega) S_I(\hbar\omega) \tag{1}$$

where the $S_I(\hbar\omega)$ corresponds to the current shot noise spectral density (Eq. (S2) in the SI) [2,3,13,16,32].

We plot the numerical results based on the 1e theory (higher order processes have not been included) together with the measured spectrum in Fig. 2a. Given the experimentally determined $\rho(\hbar\omega)$, the calculated spectrum matches the experimental data exceedingly well across the whole spectrum range for all the biases. There is a slight deviation between the theory line and the experimental data at the cutoff position for the 1.6 V curve. This can be explained by the fact that increased current at this higher bias has made the 2e process no longer negligible. This is verified as including the calculated 2e contribution results in the better match with the measured data, which is plotted in the inset in Fig. 2a. Moreover, the electron temperature in the Boltzmann factor within the 1e shot noise spectrum can be estimated via the shape of the spectrum's bias threshold in a similar way to Alberto et al.[4], with $T = 45$ K leading to acceptable fitting results (detailed fitting process documented in the section 3 of the SI). This analysis assumes that the electronic distribution in the source and drain is of the Fermi-Dirac form, with a characteristic electronic temperature. Such an electronic temperature is slightly higher than the substrate temperature $T_{sub} = 30$ K held fixed during the whole measurement. The modest amount of temperature rise, consistent with the previous light emission studies in an STM based on the coherent multielectron process[13], demonstrates some bias-driven heating of the electrons.

Earlier in the electromigration process (after the second migration step), this junction with higher conductance (now $4.8 \times 10^{-3} G_0$ zero bias conductance) exhibits a small portion of above-threshold emission beyond the 1e cutoff emission in Fig. 2b. The below-threshold portion, as before, rises expeditiously as the bias increases, while the roll off at each curve's cutoff position ($\hbar\omega = eV_b$) is much more broadened due to higher order electron tunneling and thermal smearing. The above-threshold contribution, though tiny compared to the below-threshold component, also reveals explicit LSP mode structure, which can be seen in the enlarged spectrum view in Fig. 2c. Furthermore, for a more direct comparison, we have also plotted the spectra at the same bias with different zero bias conductances together in Fig. S10. The drastic decay in the multielectron emission contribution after enlarging the gap can be clearly seen.

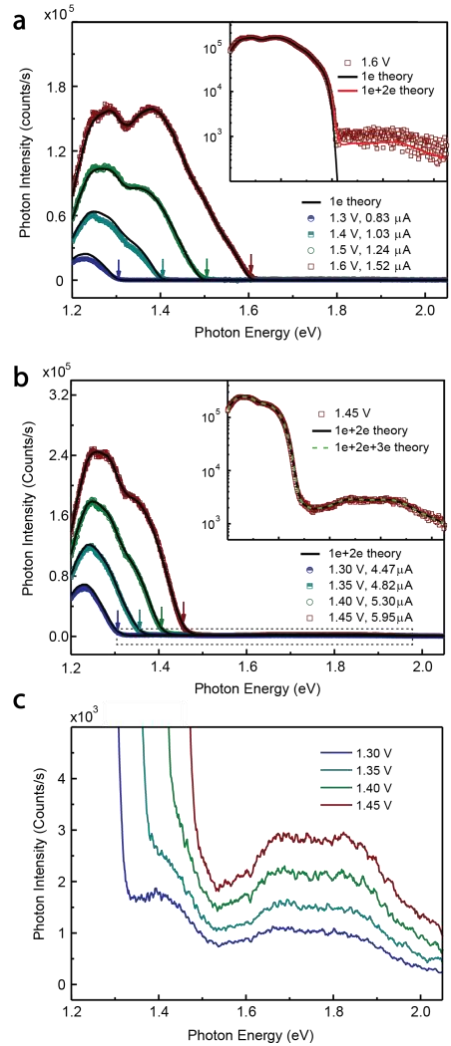

**Figure 2.** Spectral measurements for below- and above-threshold light emission at different conductance levels for the same Al tunnel junction and accompanying numerical calculation results. (a) Measured spectral light emission (colored points) after a final step of electromigration with a zero-bias conductance of $1.5 \times 10^{-3} G_0$, plotted together with numerical calculation (solid lines) for the below-threshold emission based on single electron tunneling theory (1e theory). The corresponding voltage cutoff position for each curve is indicated by the colored arrow. Tunneling current is labeled in the legend beside the applied bias for each curve. Inset shows the 1.6 V curve plotted together with the single electron (1e) and multielectron (1e+2e) tunneling theory on logarithmic scale. (b) Measured spectral light emission at an earlier stage of electromigration compared to (a) (zero-bias conductance $4.8 \times 10^{-3} G_0$), plotted together with the calculation results for multielectron tunneling (1e+2e theory). Inset shows the 1.45 V compared with multielectron theory calculated to different order (1e+2e and 1e+2e+3e). (c) Enlarged view of the spectrum region framed by the dotted line shown in (b).

The above-threshold tail, as well as the broadened abrupt change occurring at $\hbar\omega = eV_b$, is due to the multielectron coherent tunneling process no longer being negligible. This contribution can be viewed as two electrons coherently exciting a higher energy LSP to form overbias radiating photons, which can be expressed as[11–14,33]:

$$U_{2e}(V_b, \hbar\omega) = G_0^2 \rho(\hbar\omega) \int_0^{\hbar\omega} \rho(E) S_I(E) S_I(\hbar\omega - E) dE \qquad (2)$$

Hence, we can perform the same analysis as in Fig. 2a but instead substitute the higher junction conductance. As can be seen in Fig. 2b, the numerically calculated spectrum in this higher conductance regime again agrees quantitatively with the normalized experimental spectrum both for the above- and below-threshold regions, with $T = 80$ K giving the best electronic temperature estimation. The increased heating compared to the low conductance situation can be explained by increased electrical Joule heating and plasmon-related dissipative processes. The inset in Fig. 2b has also included the 3e contribution, and the tiny difference with the 1e+2e line justifies that the 3e term is negligible here.

The inelastic electron tunneling theory clearly describe the below- and above-threshold behavior as the junction conductance increases for Al junctions. Even higher conductances, however, take the junction from this regime to one where hot electron emission processes associated with higher currents dominate. As can be seen in Fig. 3a, the same Al junction with 0.09 $G_0$ zero bias starting conductance (after first step electromigration) emits a significant portion of above-threshold photons, without any obvious sharp downturns occurring at energies corresponding to the bias ($\hbar\omega = eV_b$).

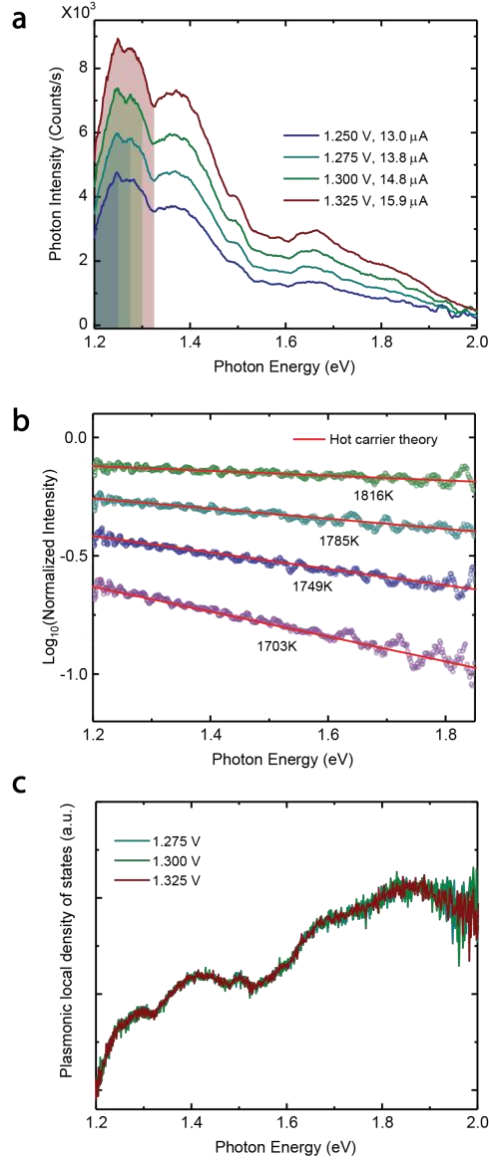

**Figure 3.** Spectral measurements for above-threshold light emission from the same tunnel junction in Fig. 2 at high current regime and the normalization analysis to examine the hot carrier distribution. (a) Measured light emission after first step of electromigration with a zero-bias conductance of $0.09G_0$. The below-threshold parts of the emission spectrum are shaded with corresponding colors of each curve. (b) Normalized spectra at different biases on logarithmic scale plotted together with fitting to the hot carrier model. The bottom line corresponds to the spectrum measured at 1.225 V. (c) Logarithmic plot of the LSP modes $\rho(\hbar\omega)$ extracted by removing the hot carrier distribution plotted in *B* from the emission spectrum in *A*. This $\rho(\hbar\omega)$ is used to obtain the numerical results for the 1e, 2e, 3e models plotted in Fig. 2.

In this high current regime where the time interval between successive tunneling electrons becomes much shorter, direct radiative decay of the electrically driven LSPs is now swamped by emission from hot electrons and hot holes above and below the Fermi energy[19,34] produced through nonradiative LSP decay. Under this scenario, the profile of the emitted spectrum is the LSP $\rho(\hbar\omega)$ coupled to the effective Boltzmann factor that describes the steady-state hot carrier contribution to the distribution, rather than the shot noise spectral density,

$$U_{hc}(V_b, \hbar\omega) \propto \rho(\omega) I^\alpha \hbar\omega \exp(-\hbar\omega/k_B T_{\text{eff}}) \qquad (3)$$

Here $\alpha$ is an exponent describing the super linear process generating the hot carriers, and is around 1.2 from analysis of prior experiments in other metals[19]. $T_{\text{eff}}$ in this model is an effective temperature characterizing the average energy of the steady-state hot carrier portion of the electronic distribution rather than describing the distribution of *all* the electrons. It is linearly proportional to the applied bias in the framework of nonradiative LSP decay as the source of hot carriers[17].

Thanks to separability between the bias dependent effective temperature in the Boltzmann factor and the energy dependent LSP modes in Eq. (3), the hot carrier distribution can be examined separately from $\rho(\hbar\omega)$ after performing a normalization analysis[19]. The linearity of the reduced spectrum plotted on logarithmic scale is obvious in Fig. 3b showing the applicability of a Boltzmann factor description with an effective temperature. We then fit the reduced spectrum with the hot carrier model, allowing inference of $T_{\text{eff}}$. These effective temperatures are comparatively lower than those extracted from pure Au junctions[19] under the same bias (~2000 K/V for the temperature bias coefficient for Au), consistent with the relative plasmonic performance for Al and Au in this photon energy range and the nonradiative plasmonic heating mechanism[35,36]. We note that this analysis is robust to the choice of which spectrum is used as the normalization reference. Reduced spectra using the 1.300 V spectrum as the reference are shown in Fig. S9, with nearly identical inferred $T_{\text{eff}}$ values. Analyzing the spectra using the obtained Boltzmann factor, a well-collapsed, energy dependent spectrum $\rho(\hbar\omega)$ is found (Fig. 3c), reflecting the plasmonic local density of states associated with the detail of the atomic-scale variance inside the electromigrated gap. This $\rho(\hbar\omega)$ is used as the input for numerically calculating the emission spectrum based on the multielectron process at low conductance regime in Fig. 2a, and 2b. The excellent consistencies between the calculated and the measured emission spectra demonstrate a robust LSP $\rho(\hbar\omega)$ throughout the whole electromigration process, indicating our controllable electromigration method selectively enlarges the interelectrode gap while still maintaining the shape of the LSP modes, despite inevitably altering their magnitude (We have also included the spectra and the numerical calculation results for another Al junction in Fig. S3 in the SI. Note that nontrivial electronic densities of states of the electrodes[7] would lead to deviations in the collapsed spectra as a function of bias.)

It is worth noting that simply by elevating the temperature in the multielectron theory, spectral intensity dependence on electrical power of the Joule heating overbias blackbody radiation can be reproduced for a high current, non-plasmonic resonant planar metallic junction[14,37], just like the Boltzmann factor. To show this, we define a total noise spectrum $\Gamma(V_b, \hbar\omega)$ which is obtained by normalizing out $\rho(\omega)$ in the overall spectrum:

$$\Gamma(V_b, \hbar\omega) = \frac{U_{total}(V_b, \hbar\omega)}{\rho(\omega)} = \frac{U_{1e}(V_b, \hbar\omega) + U_{2e}(V_b, \hbar\omega) \cdots}{\rho(\omega)} \qquad (4)$$

Parameters in Fig. 3a are used to calculate $\Gamma(V_b, \hbar\omega)$ at various temperature, as can be seen in the logarithmic plot in Fig. 4a, where the bias $V_b$ is set to be 1.325V. The abrupt turning point at $\hbar\omega = eV_b$ rapidly smears out as the temperature rises, eventually reaching an approximately linear energy dependence (~1200K) on a logarithmic scale, as seen in the Boltzmann factor. Following the calculation procedure of Fig. 2, we perform the same multielectron fit. As shown in Fig. 4b, while the fitted spectra qualitatively agree with the shape of experiment data, there are systematic deviations under the constraint of a fixed $\rho(\hbar\omega)$ and normalization. On the other hand, using the Boltzmann factor and inserting the voltage dependent temperature obtained from fitting in Fig. 3b, the calculated spectra based on hot carrier model gives extremely good reproducibility (Fig. 4c), which is expected given the fact that using the $\rho(\hbar\omega)$ all collapse together for different biases (Fig. 3c) (Detailed discussion can be found in section 9 of the SI).

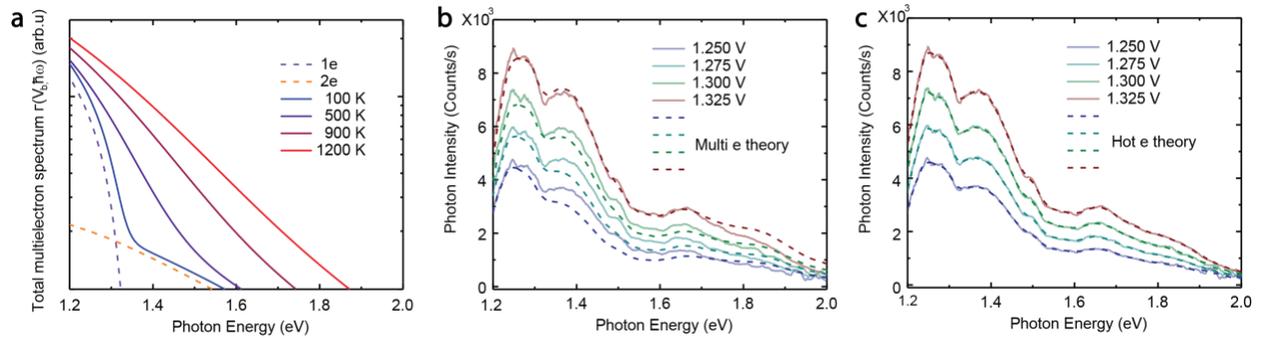

**Figure 4.** Multielectron and hot carrier calculation based on high conductance and high electronic temperature. (a) Logarithmic plot of the total noise spectrum $\Gamma(V_b, \hbar\omega)$ versus the photon energy at 1.325 V with different assumed electronic temperatures. The dotted lines are the 1e and 2e contribution for the $\Gamma(V_b, \hbar\omega)$ at an electronic temperature of 100K (b) Multielectron calculation (dotted lines) for 1.250 V, 1.275 V, 1.300 V, 1.325 V spectra using the normalization of $\rho(\hbar\omega)$ found by fitting the 1.325 V data (980K electronic temperature), and using electronic temperature as the fit parameter for each bias, plotted together with respective measured spectra in the same colors as in Fig. 3a. The fitted temperature for the spectra are 950 K(1.300 V), 920 K(1.275 V), 900 K(1.250 V) respectively. Given the constraints of the experimentally determined $\rho(\hbar\omega)$ and the conductance, the coherent multielectron model shows systematic deviations from the data with only the electronic temperature as a parameter. (c) Hot electron calculation using the effective temperature extracted from fitting $\rho(\hbar\omega)$ in Fig. 3c, showing that the calculated spectra match the data exceedingly well.

The change in dominant emission mechanism as a function of tunneling current provides important insights into the hot carrier lifetime required to maintain a steady-state hot carrier distribution in Al. Hot carriers injected through elastic tunneling or generated by non-radiative LSP damping will redistribute energy through the electronic distribution, which can be parameterized by the hot carrier lifetime on the time scale of tens to a few hundreds of fs[29,30,34,38]. On longer timescales, electron-phonon scattering will transfer energy to the lattice. The crossover from the hot carrier dominant regime to the multi-electron tunneling regime then approximately demonstrates a timescale threshold required for establishing a steady-state population of hot carriers. A rough parameter space diagram can be constructed by logarithmically plotting the photon yield, defined by the integrated photon emission rate in the visible range of spectrometer (1.12 eV band gap for silicon) divided by the tunneling rate of the electrons, versus the time interval between successive electrons, as can be seen in Fig. 4a. Different light emission regimes have been labeled into different colors, where around 8 µA (corresponding to an average time between tunneling events of around 20 fs) is approximately the crossover between having a significant steady-state hot carrier population and the multi-electron tunneling picture. This is consistent with the lifetime studies for excited electrons in Al by femtosecond time-resolved two photon photoemission[39]. Once the current falls to 2 µA (80 fs average time interval), the system is basically limited to single-electron processes, and the conventional first-order tunneling mechanism explains the data quantitatively. Within the domain of each given emission mechanism, increasing the size of the gap/decreasing the conductance lowers the photon intensity. However, different emission mechanisms have different efficiencies, so that the above-threshold hot carrier emission (Fig. 3) is less efficient per carrier than the below-threshold emission in the multi-electron inelastic tunneling mechanism. Interestingly, the multielectron process induced light emission can still be seen even at high current (~30 µA) in previous works based on the STM platform[4,13]. This is because the crossover from multielectron to hot-carrier dominated above-threshold light emission being is set by both the tunneling current and the plasmonic strength of the specific junction configuration (material, structure, geometry). In the STM-based light emission experiments, plasmonic enhancement tends to be weaker for the tip-substrate junction structure (with only a dipolar tip plasmon mode) than the planar, electromigration created metallic nanogaps (hosting both transverse and tip modes, with hybridization involving multipolar modes of the gap allow both transverse and tip modes)

It is worth noting in Fig. 5a that the photon yield per electron monotonically increases as the tunneling current decreases in the single electron tunneling regime. This observation is further strengthened through a more detailed analysis in the SI which looks at the photon yield evolution of individual junctions as the electromigration process continues, mitigating the effect of the device-to-device LSP modes variance on the overall photon yield. We further show that the intensity can be modified via plasmonic resonance engineering. As can be seen in Fig. 5b, adjusting the width of the nanowire (100 nm to 200 nm) puts the transverse plasmon resonance of the nanowire in the emission energy window, leading to a two order of magnitude photon yield enhancement at a given current level. This plasmonic effect can be quantified by the electrodynamic simulation results in the SI (section 5). This orders-of-magnitude tunability combined with the preference for the direct radiative decay and photon yield increase at the low conductance suggest the potential for Al devices in realizing efficient on-chip light sources[40–44], potentially for entangled photon generation[45].

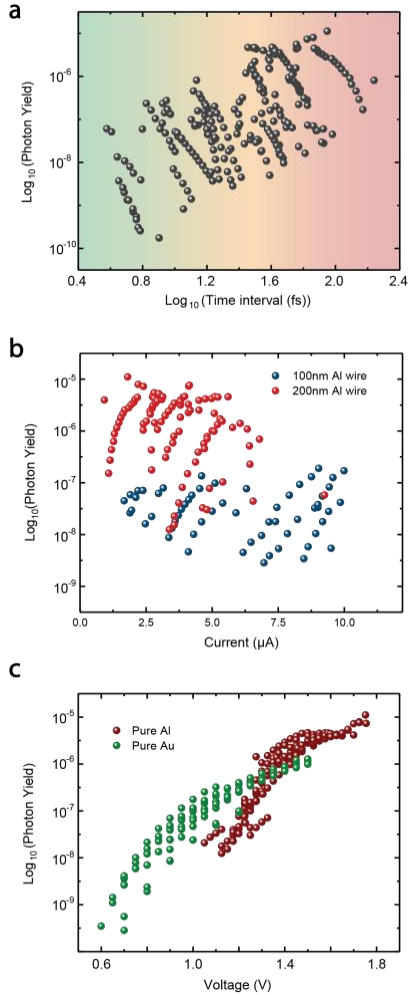

**Figure 5.** Statistical analysis for the overall photon yield. (a) Logarithmic plot of the photon yield versus the average tunneling time interval between two successive electrons for different Al junctions, which can be viewed as a crossover diagram for different light emission mechanism. The crossover behavior between different mechanisms is indicated qualitatively by the colored gradient in the background. (b) Measured photon yield in the multielectron inelastic tunneling emission regime (plotted on logarithmic scale versus tunneling current) for an ensemble of Al junctions formed by electromigrating nanowires with different widths. (c) Measured photon yield (plotted on logarithmic scale) for Al junction in the multielectron inelastic tunneling emission regime plotted together with Au junctions in the hot carrier recombination regime versus the applied bias.

Fig. 5c makes a direct comparison between photon yield for Al junctions in the multi-electron tunneling regime and pure Au junctions where the hot carrier mechanism dominates (Au data set is from our previous work[19]). The Al photon yield from the single electron tunneling below-threshold emission increases with the bias, surpassing the hot carrier photon yield from pure Au devices at around 1.4 V. The lack of $d$ electrons and interband transitions favor the direct radiative decay channel in Al, consistent with the observation that Al junction readily falls into the multi-electron cotunneling regime. In contrast, in Au based junctions, the Au dielectric function favors stronger plasmon performance in this photon energy range, where interband transitions enhance non-radiative LSP decay and comparatively favor hot carrier generation[34,35].

Our experimental results and analysis suggest that the crossover at high currents to hot carrier-based emission is most effective in systems where there are both strong plasmon resonances and interband scattering to favor non-radiative decay of electrically excited plasmon modes. Al junctions distinguish themselves clearly from the previously studied noble plasmonic materials in the field of above-threshold light emission because of aluminum's distinct plasmonic properties and electronic structure. The large imaginary part of the dielectric constant in Al enables a higher damping of local plasmons, which could potentially suppress the plasmon-based generation channel for the steady state hot carrier distribution, which is quite different from the Au junction case. It has been shown that optical excitation, when in resonance with the planar junction, is efficient in generating hot carriers, which generates photons either through direct radiative recombination[17,34] or through electronic Raman like process[46–48]. Indeed, our further experimental results (Fig. S7) show that when simultaneous optical excitation was applied to an Al tunnel junction, the above-threshold emission contribution is enhanced, while the below threshold part associated with electron inelastic tunneling processes remains unchanged.

Al junctions allow us to demonstrate that the tension regarding different light emission mechanisms in metal tunnel junctions is really a question of regimes and relative timescales. Detailed examination of light emission provides a promising approach to optimal performance for Al based junctions via specifically designed nanoconstriction geometries and fine conductance control ($\sim 1 \times 10^{-3} G_0$). These findings will further motivate using plasmonic active tunnel junctions as a tool for integrated on-chip quantum photonic devices and atomic scale light sources.

**Supporting Information**.

The following file is available free of charge on the ACS Publication website. Materials and Methods; Statistical Analysis of Electromigrated Junction Conductance; Single- and Multi-electron Inelastic Tunneling Induced light Emission: theory; Photon Yield Change During Multi-step Electromigration; Numerical Results for Al Nanowire with Different Widths; Electro-photoluminescence Compared with Electroluminescence; Heat Dissipation via the Electron-phonon Scattering.

AUTHOR INFORMATION

**Corresponding Author**

* corresponding author: Douglas Natelson Email: natelson@rice.edu

**Author Contributions**

**Funding Sources**

ONR N00014-21-1-2062, Robert A. Welch Foundation Award C-1636, and AFRL FA8651-22-1-0012.

ACKNOWLEDGMENT

D.N. and Y.Z. acknowledge ONR N00014-21-1-2062; D.N., M. A., and Y. Z. acknowledge Robert A. Welch Foundation Award C-1636; L.C. acknowledges support from AFRL FA8651-22-1-0012.

ABBREVIATIONS

LSP, localized surface plasmon.

REFERENCES

(1) Lambe, J.; McCarthy, S. L. Light Emission from Inelastic Electron Tunneling. *Phys. Rev. Lett.* **1976**, *37* (14), 923–925

(2) Kalathingal, V.; Dawson, P.; Mitra, J. Scanning Tunnelling Microscope Light Emission: Finite Temperature Current Noise and over Cut-off Emission. *Sci. Rep.* **2017**, *7* (1), 3530

(3) Zhang, C.; Hugonin, J.-P.; Coutrot, A.-L.; Sauvan, C.; Marquier, F.; Greffet, J.-J. Antenna Surface Plasmon Emission by Inelastic Tunneling. *Nat. Commun.* **2019**, *10* (1), 4949

(4) Martín-Jiménez, A.; Lauwaet, K.; Jover, Ó.; Granados, D.; Arnau, A.; Silkin, V. M.; Miranda, R.; Otero, R. Electronic Temperature and Two-Electron Processes in Overbias Plasmonic Emission from Tunnel Junctions. *Nano Lett.* **2021**, *21* (16), 7086–7092

(5) Qian, H.; Li, S.; Hsu, S. W.; Chen, C. F.; Tian, F.; Tao, A. R.; Liu, Z. Highly-Efficient Electrically-Driven Localized Surface Plasmon Source Enabled by Resonant Inelastic Electron Tunneling. *Nat. Commun. 2021 121* **2021**, *12* (1), 1–7

(6) Edelmann, K.; Wilmes, L.; Rai, V.; Gerhard, L.; Yang, L.; Wegener, M.; Repän, T.; Rockstuhl, C.; Wulfhekel, W. Influence of Co Bilayers and Trilayers on the Plasmon-Driven Light Emission from Cu(111) in a Scanning Tunneling Microscope. *Phys. Rev. B* **2020**, *101* (20), 205405

(7) Martín-Jiménez, A.; Fernández-Domínguez, A. I.; Lauwaet, K.; Granados, D.; Miranda, R.; García-Vidal, F. J.; Otero, R. Unveiling the Radiative Local Density of Optical States of a Plasmonic Nanocavity by STM. *Nat. Commun.* **2020**, *11* (1), 1021

(8) Hoffmann, G.; Berndt, R.; Johansson, P. Two-Electron Photon Emission from Metallic Quantum Wells. *Phys. Rev. Lett.* **2003**, *90* (4), 046803

(9) Schull, G.; Néel, N.; Johansson, P.; Berndt, R. Electron-Plasmon and Electron-Electron Interactions at a Single Atom Contact. *Phys. Rev. Lett.* **2009**, *102* (5), 057401

(10) Schneider, N. L.; Johansson, P.; Berndt, R. Hot Electron Cascades in the Scanning Tunneling Microscope. *Phys. Rev. B* **2013**, *87* (4), 45409

(11) Xu, F.; Holmqvist, C.; Belzig, W. Overbias Light Emission Due to Higher-Order Quantum Noise in a Tunnel Junction. *Phys. Rev. Lett.* **2014**, *113* (6), 066801

(12) Xu, F.; Holmqvist, C.; Rastelli, G.; Belzig, W. Dynamical Coulomb Blockade Theory of Plasmon-Mediated Light Emission from a Tunnel Junction. *Phys. Rev. B* **2016**, *94*, 245111

(13) Peters, P.-J.; Xu, F.; Kaasbjerg, K.; Rastelli, G.; Belzig, W.; Berndt, R. Quantum Coherent Multielectron Processes in an Atomic Scale Contact. *Phys. Rev. Lett.* **2017**, *119* (6), 066803

(14) Fung, E. D.; Venkataraman, L. Too Cool for Blackbody Radiation: Overbias Photon Emission in Ambient STM Due to Multielectron Processes. *Nano Lett.* **2020**, *20* (12), 8912–

8918

(15) Schneider, N. L.; Matino, F.; Schull, G.; Gabutti, S.; Mayor, M.; Berndt, R. Light Emission from a Double-Decker Molecule on a Metal Surface. *Phys. Rev. B* **2011**, *84*, 153403

(16) Schneider, N. L.; Schull, G.; Berndt, R. Optical Probe of Quantum Shot-Noise Reduction at a Single-Atom Contact. *Phys. Rev. Lett.* **2010**, *105* (2), 026601

(17) Shalem, G.; Erez-Cohen, O.; Mahalu, D.; Bar-Joseph, I. Light Emission in Metal–Semiconductor Tunnel Junctions: Direct Evidence for Electron Heating by Plasmon Decay. *Nano Lett.* **2021**, *21* (3), 1282–1287

(18) Ott, C.; Götzinger, S.; Weber, H. B. Thermal Origin of Light Emission in Nonresonant and Resonant Nanojunctions. *Phys. Rev. Res.* **2020**, *2* (4), 042019

(19) Cui, L.; Zhu, Y.; Abbasi, M.; Ahmadivand, A.; Gerislioglu, B.; Nordlander, P.; Natelson, D. Electrically Driven Hot-Carrier Generation and Above-Threshold Light Emission in Plasmonic Tunnel Junctions. *Nano Lett.* **2020**, *20* (8), 6067–75

(20) Pechou, R.; Coratger, R.; Ajustron, F.; Beauvillain, J. Cutoff Anomalies in Light Emitted from the Tunneling Junction of a Scanning Tunneling Microscope in Air. *Appl. Phys. Lett.* **1998**, *72* (6), 671–673

(21) Buret, M.; Uskov, A. V.; Dellinger, J.; Cazier, N.; Mennemanteuil, M.-M.; Berthelot, J.; Smetanin, I. V.; Protsenko, I. E.; Colas-des-Francs, G.; Bouhelier, A. Spontaneous Hot-Electron Light Emission from Electron-Fed Optical Antennas. *Nano Lett.* **2015**, *15* (9), 5811–5818

(22) Malinowski, T.; Klein, H. R.; Iazykov, M.; Dumas, P. Infrared Light Emission from Nano

Hot Electron Gas Created in Atomic Point Contacts. *EPL* **2016**, *114* (5), 57002

(23) Boyle, M. G.; Mitra, J.; Dawson, P. The Tip–Sample Water Bridge and Light Emission from Scanning Tunnelling Microscopy. *Nanotechnology* **2009**, *20* (33), 335202

(24) Knight, M. W.; King, N. S.; Liu, L.; Everitt, H. O.; Nordlander, P.; Halas, N. J. Aluminum for Plasmonics. *ACS Nano* **2014**, *8* (1), 834–840

(25) Knight, M. W.; Liu, L.; Wang, Y.; Brown, L.; Mukherjee, S.; King, N. S.; Everitt, H. O.; Nordlander, P.; Halas, N. J. Aluminum Plasmonic Nanoantennas. *Nano Lett.* **2012**, *12* (11), 6000–6004

(26) Lopez-Acevedo, O.; Clayborne, P. A.; Häkkinen, H. Electronic Structure of Gold, Aluminum, and Gallium Superatom Complexes. *Phys. Rev. B* **2011**, *84* (3), 035434

(27) Zhu, Y.; Cui, L.; Natelson, D. Hot-Carrier Enhanced Light Emission: The Origin of above-Threshold Photons from Electrically Driven Plasmonic Tunnel Junctions. *J. Appl. Phys.* **2020**, *128* (23), 233105

(28) Ward, D. R.; Hüser, F.; Pauly, F.; Cuevas, J. C.; Natelson, D. Optical Rectification and Field Enhancement in a Plasmonic Nanogap. *Nat. Nanotechnol.* **2010**, *5* (10), 732–736

(29) Liu, J. G.; Zhang, H.; Link, S.; Nordlander, P. Relaxation of Plasmon-Induced Hot Carriers. *ACS Photonics* **2018**, *5* (7), 2584–2595

(30) Sun, C.-K.; Vallee, F.; Acioli, L. H.; Ippen, E. P.; Fujimoto, J. G. Femtosecond-Tunable Measurement of Electron Thermalization in Gold. *Phys. Rev.* **1994**, *50* (20), 15337–15348

(31) Herzog, J. B.; Knight, M. W.; Li, Y.; Evans, K. M.; Halas, N. J.; Natelson, D. Dark

Plasmons in Hot Spot Generation and Polarization in Interelectrode Nanoscale Junctions. *Nano Lett.* **2013**, *13* (3), 1359–1364

(32) Paoletta, A. L.; Fung, E. D.; Venkataraman, L. Gap Size-Dependent Plasmonic Enhancement in Electroluminescent Tunnel Junctions. *ACS Photonics* **2022**, *9* (2), 688–693

(33) Kaasbjerg, K.; Nitzan, A. Theory of Light Emission from Quantum Noise in Plasmonic Contacts: Above-Threshold Emission from Higher-Order Electron-Plasmon Scattering. *Phys. Rev. Lett.* **2015**, *114* (12), 126803

(34) Cui, L.; Zhu, Y.; Nordlander, P.; Ventra, M. Di; Natelson, D. Thousand-Fold Increase in Plasmonic Light Emission via Combined Electronic and Optical Excitations. *Nano Lett.* **2021**, *21* (6), 2658–2665

(35) Sundararaman, R.; Narang, P.; Jermyn, A. S.; Goddard, W. A.; Atwater, H. A. Theoretical Predictions for Hot-Carrier Generation from Surface Plasmon Decay. *Nat. Commun.* **2014**, *5* (1), 1–8

(36) Narang, P.; Sundararaman, R.; Atwater, H. A. Plasmonic Hot Carrier Dynamics in Solid-State and Chemical Systems for Energy Conversion. *Nanophotonics* **2016**, *5* (1), 96–111

(37) Downes, A.; Dumas, P.; Welland, M. E. Measurement of High Electron Temperatures in Single Atom Metal Point Contacts by Light Emission. *Appl. Phys. Lett.* **2002**, *81* (7), 1252–1254

(38) Su, M. N.; Ciccarino, C. J.; Kumar, S.; Dongare, P. D.; Hosseini Jebeli, S. A.; Renard, D.; Zhang, Y.; Ostovar, B.; Chang, W. S.; Nordlander, P.; et al. Ultrafast Electron Dynamics in Single Aluminum Nanostructures. *Nano Lett.* **2019**, *19* (5), 3091–3097

(39) Bauer, M.; Pawlik, S.; Aeschlimann, M. Electron Dynamics of Aluminum Investigated by Means of Time-Resolved Photoemission. *Laser Tech. Surf. Sci. III* **1998**, *3272*, 201–210

(40) Qian, H.; Hsu, S.-W.; Gurunatha, K.; Riley, C. T.; Zhao, J.; Lu, D.; Tao, A. R.; Liu, Z. Efficient Light Generation from Enhanced Inelastic Electron Tunnelling. *Nat. Photonics* **2018**, *12* (8), 485–488

(41) Du, W.; Wang, T.; Chu, H.-S.; Nijhuis, C. A. Highly Efficient On-Chip Direct Electronic–Plasmonic Transducers. *Nat. Photonics* **2017**, *11* (10), 623–627

(42) Kern, J.; Kullock, R.; Prangsma, J.; Emmerling, M.; Kamp, M.; Hecht, B. Electrically Driven Optical Antennas. *Nat. Photonics* **2015**, *9* (9), 582–586

(43) Parzefall, M.; Szabó, Á.; Taniguchi, T.; Watanabe, K.; Luisier, M.; Novotny, L. Light from van Der Waals Quantum Tunneling Devices. *Nat. Commun.* **2019**, *10* (1), 1–9

(44) Parzefall, M.; Bharadwaj, P.; Jain, A.; Taniguchi, T.; Watanabe, K.; Novotny, L. Antenna-Coupled Photon Emission from Hexagonal Boron Nitride Tunnel Junctions. *Nat. Nanotechnol.* **2015**, *10* (12), 1058–1063

(45) Leon, C. C.; Rosławska, A.; Grewal, A.; Gunnarsson, O.; Kuhnke, K.; Kern, K. Photon Superbunching from a Generic Tunnel Junction. *Sci. Adv.* **2019**, *5* (5), eaav4986

(46) Hugall, J. T.; Baumberg, J. J. Demonstrating Photoluminescence from Au Is Electronic Inelastic Light Scattering of a Plasmonic Metal: The Origin of SERS Backgrounds. *Nano Lett.* **2015**, *15* (4), 2600–2604

(47) Mertens, J.; Kleemann, M.-E.; Chikkaraddy, R.; Narang, P.; Baumberg, J. J. How Light Is Emitted by Plasmonic Metals. *Nano Lett.* **2017**, *17* (4), 2568–2574

(48) Dey, S.; Banik, M.; Hulkko, E.; Rodriguez, K.; Apkarian, V. A.; Galperin, M.; Nitzan, A. Observation and Analysis of Fano-like Lineshapes in the Raman Spectra of Molecules Adsorbed at Metal Interfaces. *Phys. Rev. B* **2016**, *93* (3), 035411

# Supporting Information

# Tuning light emission crossovers in atomic-scale aluminum plasmonic tunnel junctions

*Yunxuan Zhu[1], Longji Cui[2,3], Mahdiyeh Abbasi[4] and Douglas Natelson[1,4,5,*]*

[1]Department of Physics and Astronomy, Rice University, Houston, TX 77005, United States

[2]Paul M. Rady Department of Mechanical Engineering, University of Colorado, Boulder, CO 80309, United States.

[3]Materials Science and Engineering Program, University of Colorado, Boulder, CO 80309, United States.

[4]Department of Electrical and Computer Engineering, Rice University, Houston, TX 77005, United States.

[5]Department of Materials Science and Nanoengineering, Rice University, Houston, TX 77005, United States.

[*]Corresponding author: Douglas Natelson (natelson@rice.edu.)

**Supplementary Information Text**

**1. Materials and Methods**

We measure the emission spectra of Al tunnel junctions in an ultrahigh vacuum optical cryostat at a substrate temperature of 30 K. Arrays of planar nanowires (600nm long with 100 nm or 200 nm width) together with the bow-tie shaped extended electrodes (as can be seen in the sketch setup for the experiment in Fig. S1a) are defined by standard e-beam lithography techniques. Two layers of e-beam resist (PMMA 495 and PMMA 950) are spin coated consecutively prior to the lithography to ease the lift off process of e-beam evaporated pure Al metal film (18 nm thick) without any adhesion layer. Multi-step electromigration [1–4] is employed to break each nanowire and form a subnanometer sized gap in a relatively controlled way. In brief, we apply a voltage sweep from 0 V to 1 V and concurrently measure the resistance change of the nanowire, until the resistance drops slightly below 1 $G_0 = 2e^2/h$. The initial electrical characterization of a nanowire after the first step of electromigration is shown in Fig. S1b. We then monotonically increase the DC bias while adding a small AC amplitude (25 mV) at low frequency (5 Hz). The tunnelling gap is then enlarged in a slow and relatively controllable manner to a desirable conductance range for data acquisition. By this approach, light emission from the same electromigrated Al junction can be measured at different gap sizes, showing the transition from above- to below-threshold emission as the tunneling current at a given voltage bias decreases. Another way of slowing down the electromigration process is also feasible by decreasing bias the before the sudden rupture of the nanowire to maintain the time evolution of the conductance[5]. Photons emitted from the biased junction are collected through an objective and directed to a CCD spectrometer. We measured emission from a total of 47 Al junctions in this work.

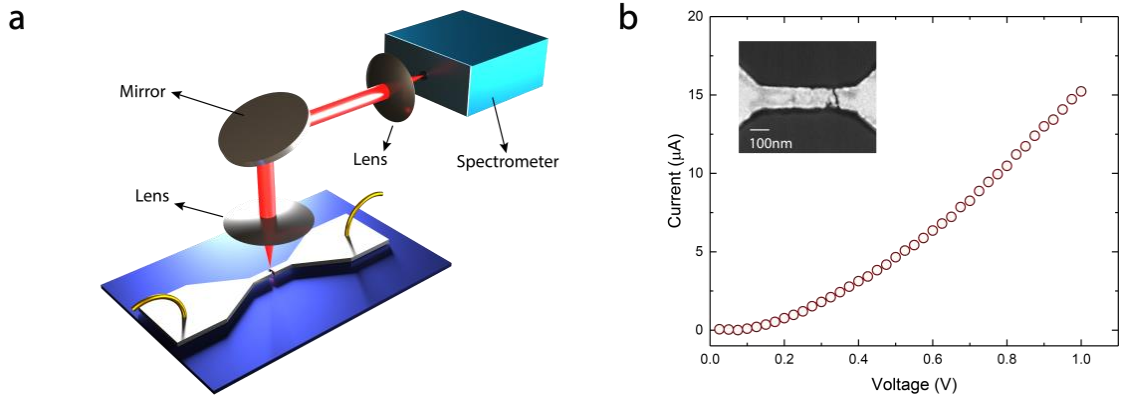

**Figure S1.** Experimental setup and initial electrical characterization of the electromigrated junction. (a) Sketch of the experimental setup. (b) I-V characteristic of a typical tunnel junction ($0.13G_0$ zero-bias conductance) after first step of electromigration. Inset shows an SEM image for an electromigrated junction.

## 2. Statistical Analysis of Electromigrated Junction Conductance.

To show the different behaviors during multi-step electromigration of tunnel junctions made of different materials, we plot the histogram of the zero-bias electrical conductance for three types of junctions (pure Au, Au with a thin Cr adhesion layer, and pure Al junctions). As shown in Fig. S2, Au and Au/Cr junctions behave similarly after the first and second step of electromigration, where the value of the conductance lies between 0.1 and 1 $G_0$. However, we note that for Au and Au/Cr junctions, it is challenging to further enlarge the tunneling gap after two steps of electromigration and the process turns uncontrollable and typically leads to a tunnel junction with a very low conductance ($< 10^{-3}G_0$) with negligible light emission. In contrast, pure Al junctions can be electromigrated step by step in controllable manner, giving rise to tunnel junctions with conductance down to the $10^{-3}G_0$ regime where the single and multi-electron inelastic tunneling processes works as the dominant light emission mechanism. Therefore, due to the good malleability, Al is a more suitable material for tuning through different conductances and emission regimes.

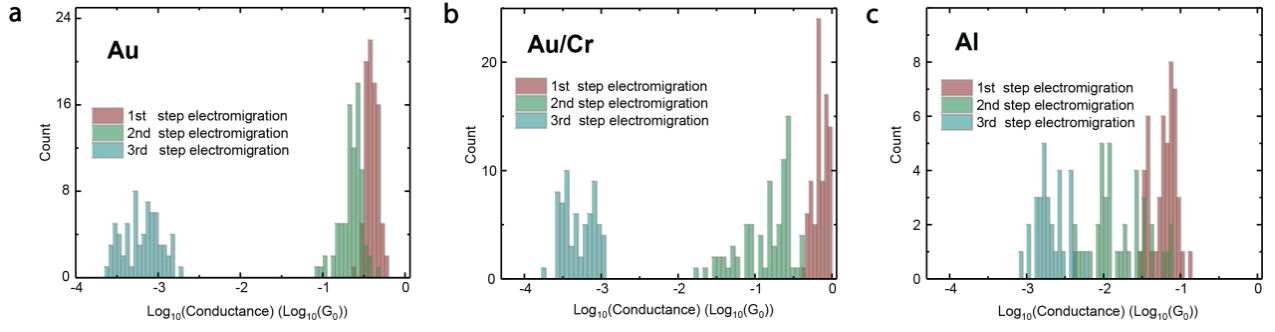

**Figure S2**. Statistical analysis for zero-bias junction conductance. (a-c) Histograms of junction conductance after successive steps of electromigration for pure Au, Au/Cr, and pure Al junctions, respectively.

## 3. Single- and Multi-electron Inelastic Tunneling Induced Light Emission: Theory

### 3.1 Single electron light emission model

In the low current limit, the electrically driven localized surface plasmon (LSP) that results in the far field light emission is excited by individual electron inelastic tunneling events thanks to the prolonged time interval between successive tunneling electrons. Using this physical picture, inelastic electron tunneling through the tunneling current fluctuations and expressed via shot noise spectral density. [6–8] The radiative decay of the electrically excited LSPs which leads to photon emission can then be modelled as the interaction between plasmonically modified local density of states ($\rho(\hbar\omega)$) and the shot noise spectrum ($S_I(\hbar\omega)$), [6–8]

$$U_{1e}(V_b, \hbar\omega) = \rho(\hbar\omega) S_I(\hbar\omega) \tag{S1}$$

where $U_{1e}$ denotes that Eq. (1) in the main text is only for the single electron tunneling process. $S_I(\hbar\omega)$ is given by [9–11]

$$S_I(V_b, \hbar\omega) = G_0 \left[ 2 \sum_n \tau_n^2 H(\hbar\omega) + \sum_n \tau_n(1-\tau_n) \sum_{\pm} H(\hbar\omega \pm eV_b) \right] \tag{S2}$$

where the summation over all the subscript $n$ considers contributions from all noninteracting conductance channels with transmission probability $\tau_n$, which can be extracted from the zero-bias conductance given by $G = \sum_n \tau_n G_0$. $H(x) = x n_B(x)$, where $n_B(x) = (exp(x/k_B T) - 1)^{-1}$ is the Bose-Einstein distribution. Since here we are in the very low conductance regime ($\sim 10^{-3} G_0$), hence we can safely assume for Al that only one conductance channel is contributing to the tunneling current with low transmission coefficients and ignore all higher order transmission terms.

**3.2 Multi-electron tunneling light emission model**

The physical mechanism of single electron inelastic tunneling induced light emission is valid for low-current tunnel junctions. As the tunneling current increases, multi-electron contributions become larger with the shortened time interval between successive tunneling electrons. The coherent multi-electron tunneling excites higher energy LSPs, which subsequently decay radiatively into emission of above-threshold photons. Eq. (S1) above only considers single-electron contribution, which contains a small portion of the overbiased emission due to finite temperature smearing effect near $\hbar\omega = eV_b$ (photons possessing excess energy on the order of $\sim k_B T$). Multi-electron effects can be modeled based on Eq. (S1) by introducing high-order interactions (mathematically, the convolution between $S_I(\hbar\omega)$ and $\rho(\hbar\omega)$). The second order contribution, which represents the two-electron coherent interaction induced light emission, can be expressed as[8,11–13]

$$U_{2e}(V_b, \hbar\omega) = G_0^2 \rho(\hbar\omega) \int_0^{\hbar\omega} \rho(E) S_I(E) S_I(\hbar\omega - E) dE \quad (S3)$$

Note here that the upper limits of the integration have been adjusted to include the second order contribution across the whole spectral range and to extend the low temperature approximation[11].

Similarly, the three-electron process $U_{3e}(V_b, \hbar\omega)$ can be obtained via generalization of Eq. (S3), and is written as[8]

$$U_{3e}(V_b, \hbar\omega) = G_0^3 \rho(\hbar\omega) \int_0^{\hbar\omega} \rho(E) S_I(E) dE \int_0^{\hbar\omega - E} \rho(E') S_I(E') S_I(\hbar\omega - E - E') dE' \quad (S4)$$

Therefore, the total light emission can be represented by summing the contribution from the single electron process and higher order effects,

$$U_{total}(V_b, \hbar\omega) = U_{1e}(V_b, \hbar\omega) + U_{2e}(V_b, \hbar\omega) + U_{3e}(V_b, \hbar\omega) \cdots \quad (S5)$$

**3.3 Theoretical calculation of $U_{total}(V_b, \hbar\omega)$**

To calculate $U_{total}(V_b, \hbar\omega)$, we need to know the zero-bias junction conductance which can be measured experimentally, and $\rho(\hbar\omega)$. Due to the rapid roll-off of light emission spectra near the cutoff energy $\hbar\omega = eV_b$ for the shot noise spectral intensity (Eq. (S2)), it is challenging to obtain $\rho(\hbar\omega)$ across the whole spectrum from only the measured data in the single/multielectron inelastic tunneling regime. However, $\rho(\hbar\omega)$ can be inferred in the hot carrier recombination regime, as the Boltzmann distribution factor does not significantly suppress the high energy portion of the spectrum, as can be seen in Fig. 3a in the main text. Following a normalization analysis procedure we applied in the previous work[3], we can extract $\rho(\hbar\omega)$, as shown in Fig. 3c. The extracted $\rho(\hbar\omega)$ is then used as input for calculating the inelastic tunneling induced spectrum. We note that using $\rho(\hbar\omega)$ in the high current limit as the substitute for the low conductance $\rho(\hbar\omega)$ requires that the shape $\rho(\hbar\omega)$ remains unchanged for the same tunnel junction. We have performed the same analysis on different devices during several steps of electromigration, showing that our multi-step electromigration does not appreciably change $\rho(\hbar\omega)$ of the same junction (as can be seen in Fig. S3 for another device). Despite the good consistency with the numerical results, the spectra exhibit a different shape from the spectra in Fig. 2. This originates from the random asymmetric geometries inside the nanogap formed during the electromigration process, which will significantly influence the hybridization of localized surface plasmons. Furthermore, while the junction-to-junction plasmon spectrum shape variance is inevitable, characteristic peak energies in the numerically calculated plasmonic radiative field enhancement (Fig. S6) for the nanowire with same aspect ratio are always preserved in the measured spectrum for different Al junctions (~1.4 eV and ~1.65 eV respectively). These peaks are highly dependent on the specific material and the general geometry (such as width, length and thickness) of the junction. In general, the specific material choice and general geometry constrain the junction spectra, with each junction having a different spectral shape because nanoscale details of the gap geometry affect LSP hybridization, while subsequent gentle migration of an existing junction does not alter the shape of its $\rho(\omega)$ appreciably.

Another approach to obtain the $\rho(\hbar\omega)$ is to further electromigrated the junction and apply a much higher bias so that $\rho(\hbar\omega)$ at higher energy is not suppressed[8,14]. However, limited by the resolution of our CCD, light emission spectrum at very small current is hard to measure.

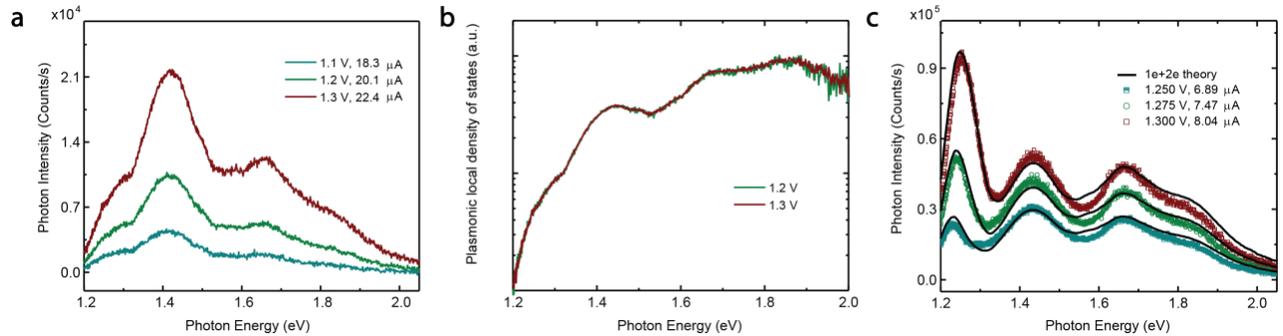

**Figure S3**. Same analysis for the light emission for another device after multi-step electromigration. (a) Measured light emission after first step of electromigration with a zero-bias junction conductance of $0.11 G_0$. (b) Logarithmic plot of extracted $\rho(\hbar\omega)$ based the spectrum in $A$. (c) Numerical calculation of the emission spectrum based on the multi-electron inelastic tunneling theory plotted together with the measured spectrum for the same device after a further step of electromigration (zero-bias conductance of $0.03 G_0$)

After inserting $\rho(\hbar\omega)$ and the zero-bias conductance, the temperature inside the shot noise spectrum in Eq. (S2) and a prefactor of the spectrum are used as the only fitting parameters to fit with the measured highest bias curve. Subsequently, the prefactor and the temperature were kept fixed to reproduce other curves with different biases in Fig. 2a and Fig. 2b in the main text.

The above-mentioned fitting can also provide a way to extract the temperature inside Eq. (S2) for the biased junction. As can be seen in Fig. S4, different temperatures have been used to reproduce the spectrum to compare with the measured spectrum at 1.6 V. As the temperature rises, the sharp downturn at the cutoff threshold broadens rapidly, eventually surpassing the measured data above $T = 45$K. The rapid evolution of the calculated spectrum with $T$ enables an estimation of the junction temperature at an accuracy within 20K.

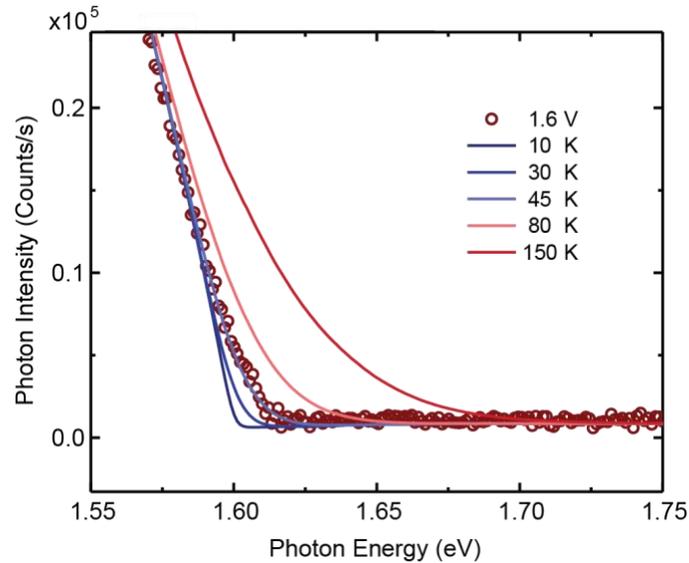

**Figure S4**. Numerical calculation results for emission from multielectron inelastic tunneling (1e+2e theory) assuming different electron temperatures, plotted together with the measured 1.6 V spectrum.

## 4. Photon Yield Change During Multi-step Electromigration

In order to eliminate the effect of the device dependent $\rho(\hbar\omega)$ on the photon yield analysis during multistep electromigration, we have calculated the ratio of the photon yield for the same junction with different zero bias conductances (after different steps of electromigration) at the same bias and plot different devices together in Fig. S5 with different symbols. The ratio is between photon yield after first electromigration step and photon yield after succeeding electromigration to a lower conductance. It can be clearly seen in Fig. S5 that all of the photon yield ratios for the same junction across different conductances configuration is larger than 1, indicating that the photon yield for one junction monotonically increases as the conductance continues to decrease.

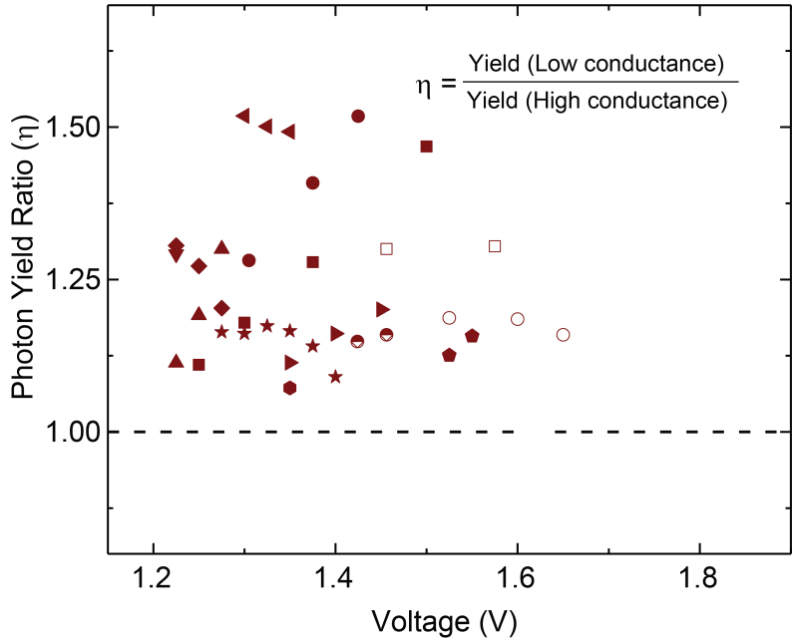

**Figure S5.** Scatter plot of photon yield ratio between the low conductance regime to the high conductance regime for the same device and same bias. Different symbols represent different devices.

## 5. Numerical Results for Al Nanowire with Different Widths

For a better understanding of the photon yield difference between the 100nm wide and 200nm wide Al nanowires, we have used representative finite-element modeling to calculate the plasmon-induced radiative enhancement for those nanowires. Previously, we have shown that the plasmonic behavior of the nanogap in the middle of the metallic nanowire is directly tied to the plasmonic characteristics of the nanowire itself[3,15]. Here we model the absorption cross section of the Al nanowire when a plane wave at different wavelength (1.2eV-2.1eV) is applied with polarization perpendicular to the width of the nanowire. Energy dependent permittivity of the Al nanowire is implemented using the algorithm for the determination of intrinsic optical constants in Al[16]. The optical absorption cross section at each step is simulated for Al nanowires with widths of 100nm and 200nm. The results are shown below. As can be seen in Fig. S6, it is shown that the normalized absorption cross section of the Al nanowire with 200nm width is much larger. This proves that the strong coupling of the incoming light with the 200nm wide Al nanowire due to activation of plasmons, consistent with the photon yield data plotted in Fig. 5c.

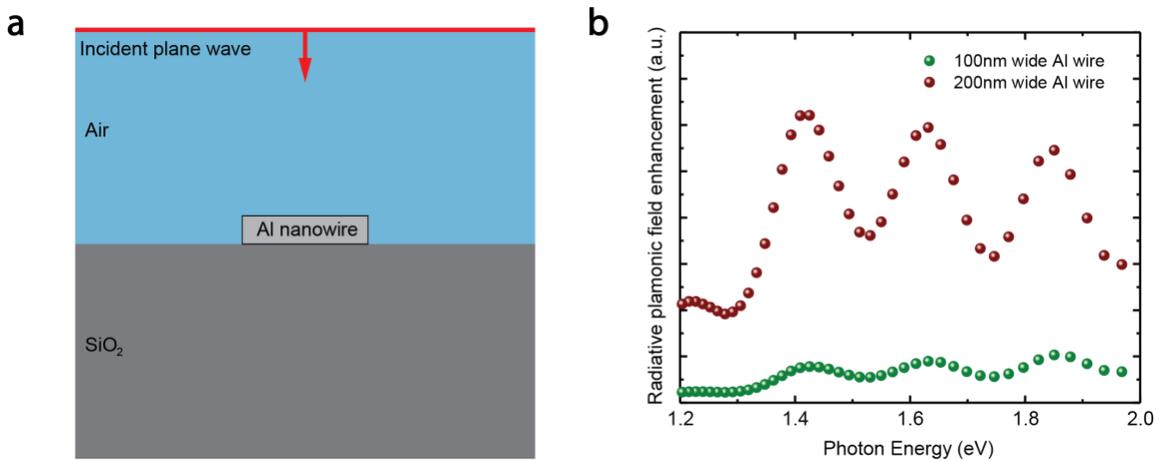

**Figure S6.** Numerically calculated plasmonic radiative filed enhancement. (a) Cross section for the simulated geometry. Al nanowire has a thickness of 18nm. (b) Calculated plasmonic radiative filed enhancement for 100nm and 200nm aluminum nanowires, plotted on linear scale.

## 6. Electro-photoluminescence Compared with Electroluminescence

To further confirm the hot carrier origin for the Al tunnel junction in the high current limit, we have performed electro-photoluminescence measurements where both electrical and optical excitation (785 nm wavelength, optical power 120 µW, focused onto the junction area with spot size 1.8 µm) are present on an Al junction in the intermediate conductance regime where the hot carrier recombination and multielectron inelastic tunneling coexist. As can be seen in Fig.S7, the above-threshold part of the emission spectrum is enhanced in the presence of optical excitation, consistent with our observation for the hot carrier enhanced electrophotoluminescence[2]. The below threshold part associated with electron inelastic tunneling processes remains unchanged, indicating that two different emission mechanisms coexist for this junction.

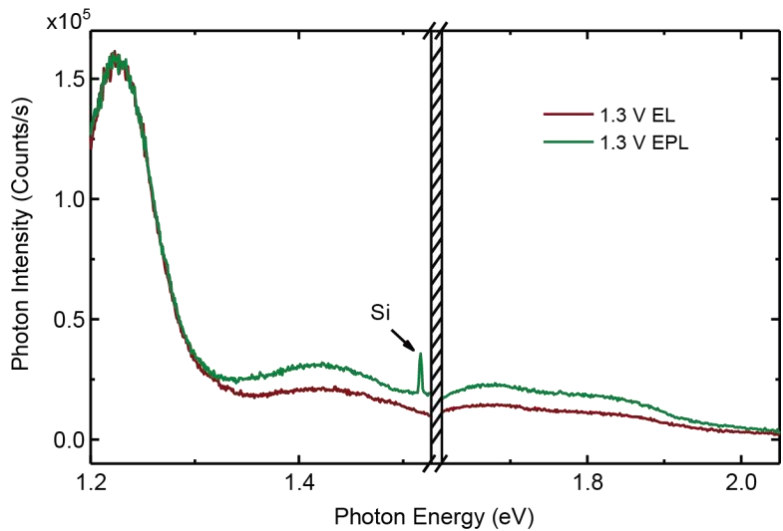

**Figure S7.** Electroluminescence (EL) at 1.3 V plotted together with the electro-photoluminescence (EPL, when simultaneous electrical and optical excitations are applied). The peak in the EPL spectrum is the Raman 520 cm$^{-1}$ mode of the silicon substrate. This Al junction has a zero-bias conductance $0.031G_0$.

## 7. Heat Dissipation via the Electron-phonon Scattering

In addition to the heat dissipation via electron-electron scattering (establishing the steady state hot carrier distribution), the local heating can also be dissipated through electron-phonon scattering to heat the lattice (corresponding to the real temperature in the shot noise spectrum). Due to the randomness of the electromigration, some uncontrollable atomic scale asymmetries can formed within the gap. However, the heat dissipation from the electron to lattice phonon is mainly determined by the geometry of the nanowire (the length and width)[17], and thus are similar for different tunnel junctions with the same nanowire aspect ratio. This is consistent with our similar fitted temperature values for different junction in the single or multielectron emission regime.

## 8. Gap distance estimation through the Simmons model

The *I-V* characteristics of the electromigrated Al tunnel junctions are measured and approximated using the Simmons model to estimate the size of the tunneling gap[18]. As shown in Fig. S8, the *I-V* curves are measured for the same Al junction after three steps of consecutive electromigration. The zero-bias conductance of the tunnel junction is decreased from $0.09G_0$ to $1.5 \times 10^{-3}G_0$. After applying the Simmons model to fit the measured *I-V* curves it is found that the tunneling gap is widened from ~ 0.17 nm to ~ 0.24 nm for a tunneling barrier height assumed to be 2.5 eV. A range of barrier heights from 1.5 eV to 5 eV has also been applied to fit the model, resulting in the estimated tunneling gap size from ~ 0.1 nm to ~ 0.4 nm.

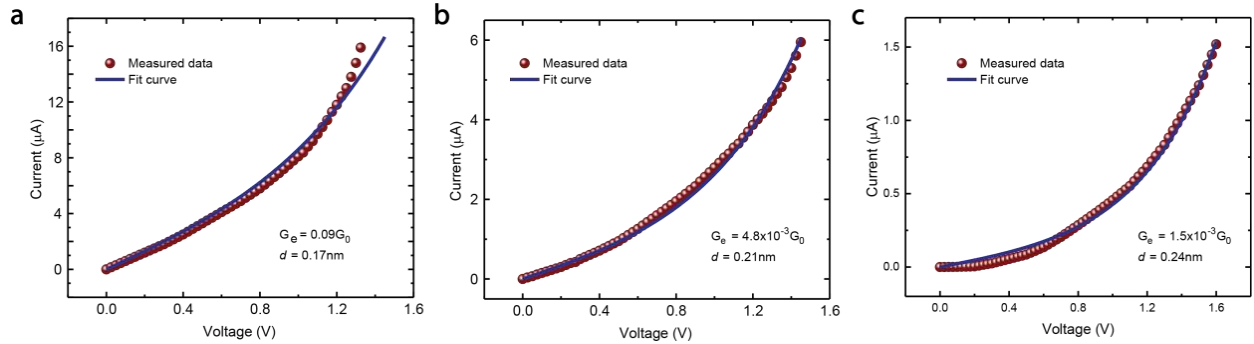

**Figure S8.** I-V fitting based on the Simmons model for the same junction with different zero bias conductance. (**a-c**) Measured I-V curve plotted together with Simons model fitting after three steps of consecutive electromigration. Gap distances obtained from the fitting are indicated in the legend.

## 9. Multielectron model at high temperature limit

It is worth considering whether the coherent multielectron model (Eq. (2) and higher order contributions) can be used to reproduce the above-threshold emission spectrum in Fig. 3a. Indeed, it has been shown previously that simply by elevating the temperature in the multielectron theory, spectral intensity dependence on electrical power of the Joule heating overbias blackbody radiation can be reproduced for a high current, non-plasmonic resonant planar metallic junction[11,19]. This can be explained by the qualitatively similar photon energy dependence with the Boltzmann factor when a much higher temperature is used. This is addressed in Fig. 4 of the main manuscript.

By elevating their respective temperature parameters, both the multielectron and hot carrier model can be applied to the above-threshold emission in the high current regime. Discussing the meaning of the temperature parameters in these models gives insight. In the multielectron tunneling case, elevating the temperature used in the current noise means that the whole electron Fermi sea (from which the transporting carriers are drawn) is heated, resulting in a very broadened electronic distribution. It is unclear what mechanism could broaden the electronic distribution for the entire electron system. In the hot carrier model, either through direct Joule heating[17,20,21] or plasmonic mediated heating[3,22,23], the temperature parameter characterizes a steady state hot carrier contribution above the remaining Fermi sea, only involves a small portion of the electrons. Note that an effective electronic temperature of ~1800 K (~150 meV) implies that to produce the above-threshold photons (energy ~10 $k_B T$), only a small fraction of the hot carriers can effectively participate (~$10^{-5}$). This is consistent with the overall photon yield decreasing when the hot carrier recombination mechanism starts to dominate in the high current regime.

To further discriminate between these models when the tunneling rates are high, both spatial- and time-resolved experimental techniques are required, given that the major difference between the models are the number of electrons involved in the effectively hot electronic distribution. In the multielectron case, a highly broadened Fermi-Dirac distribution implies a much larger energy density of heated carriers at the nanoscale, and therefore would be expected to have distinct photon emission features at the extreme near field. Detailed experimental characterization of such emission properties is beyond the scope of this work. Here, the $\rho(\hbar\omega)$ obtained in Fig. 3c leading to excellent agreement with the low current spectra via multielectron theory indicates that hot carrier model gives a more straightforward and accurate description at high current limit.

Theoretical analysis for light emission in the hot carrier regime is performed following the detailed procedure described in previous work[3]. Briefly, we can obtain the plasmonic resonance through normalization analysis which extracts the effective temperature of the hot carriers from the Boltzmann factor. As shown in Fig. 3b, the collapse of voltage independent plasmonic resonance and the linear form of the logarithmic normalized spectrum validate the plasmonic driven hot carrier model.

We note that the selection of the particular light emission reference curve used in the normalization analysis does not influence the inferred electronic temperatures, further supporting the conclusions and validation of the theoretical model. For example, choosing the reference spectrum measured at 1.300 V rather than the one measured at 1.325 V (as in Fig. 3 of the main text) as the normalization reference, all the inferred temperatures at different biases differ from the 1.325 V reference case by only a few Kelvins, well within the fitting error (~50K) for the Boltzmann factor, as shown by comparing Fig. S9 with Fig. 3b.

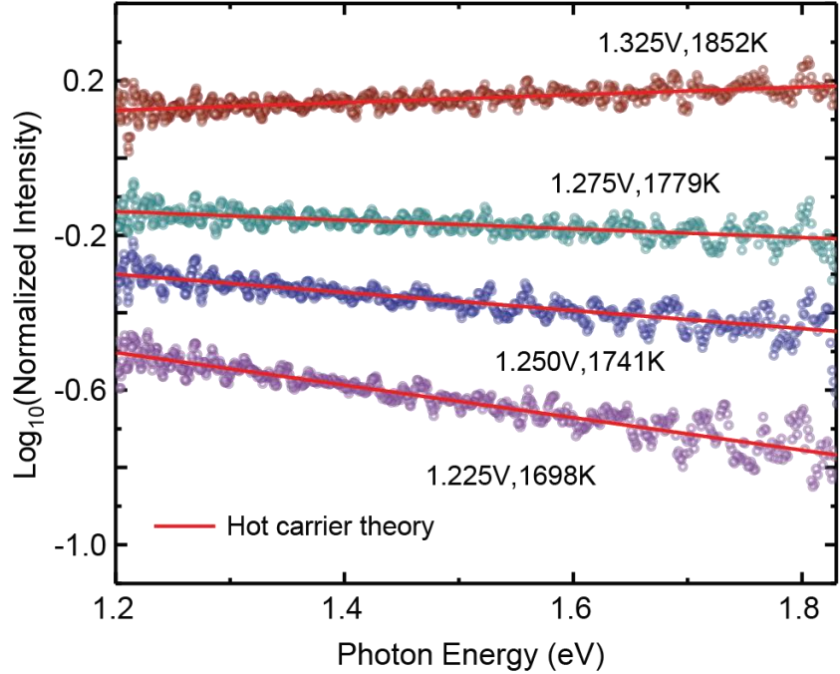

**Figure S9.** Normalized spectra at different biases on logarithmic scale plotted together with fitting to the hot carrier model. The data set here is identical to that of Fig. 3 in the main text, while here the normalization reference curve is chosen to be the spectrum measured at 1.300 V.

## 10. Comparison between the STM break junctions

Based on the multi-step electromigration method[2,3], we are able to tune the dominant mechanism of light emission from the Al junction systematically as the tunneling conductance is decreased with each electromigration step. This experimental strategy is very similar to actively controlling the gap surface distance in an STM experiment[11,24], though in an irreversible way due to the nature of electromigration. In the STM break junction, solely increasing the current will drive the system from overbias emission solely due to $k_B T$ smearing of the single-electron process[25] to a multielectron dominant above-threshold emission, eventually reaching an intermediate regime where the multielectron tunneling and hot carrier emission coexist[14]. As we have noted in the main text, the plasmonic structure of planar junctions involves both tip and transverse plasmons, with hybridization involving highly localized multipolar modes. In contrast, in the STM geometry, typically only the dipolar tip mode is relevant. One way to test the importance of plasmonic mode structure to the emission mechanism crossover would be designing active plasmonic structures underneath the STM tip. Another way to introduce the hot carrier distribution is leveraging optical excitation.

Comparison between the STM and those planar tunnel junctions can provide further insights. The Angstrom scale continuous tunability of an STM junction makes it obviously more controlled than the random junction gap created by electromigration. The ability to map out emission intensity as a function of tip location against different substrate positions provides additional information regarding plasmonic/electronic/structural details of the substrate. The STM technique can also operate in the feedback-controlled scanning mode, which enables the imaging of the delicate light emission property of a plasmonic nanostructure, which is lacking in the planar electromigration junction experiment. In contrast, the planar junction is more amenable to conventional optical microscopy geometries and can be flexibly scaled up into an array structure using conventional top-down fabrication techniques, which facilitate on-chip massive designs and applications. In addition, the planar structure enables integration with gate electrodes, which is challenging to realize in a typical STM platform. Such tunability may enable future directions.

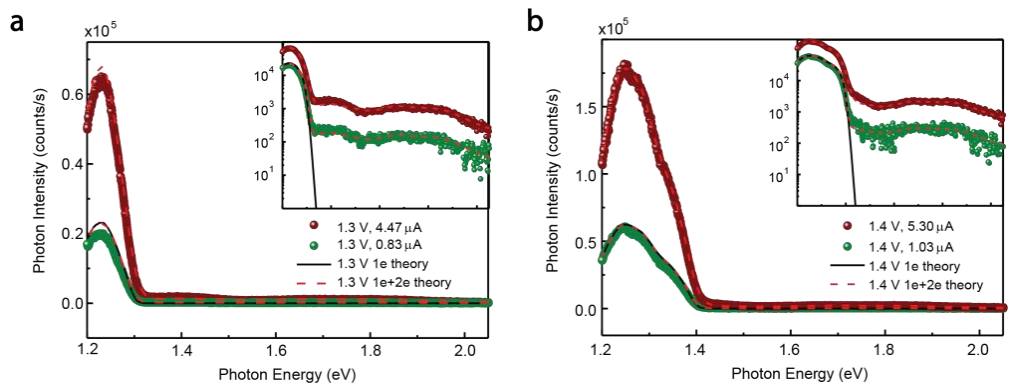

**Figure S10.** Spectral measurements for light emission at same biases with different conductance levels for the same Al tunnel junction (Fig. 2) and accompanying numerical calculation results. (a) Measured spectral light emission (colored points) with a zero-bias conductance of $4.8 \times 10^{-3} G_0$, plotted together with numerical calculation (solid lines) based on single electron tunneling theory (1e theory) and multielectron theory (1e+2e theory). Inset shows the same plot on logarithmic scale. (b) Measured spectral light emission (colored points) with a zero-bias conductance of $1.5 \times 10^{-3} G_0$ and numerical calculation (solid lines) results.

**Supporting Information References**

(1) Park, H.; Lim, A. K. L.; Alivisatos, A. P.; Park, J.; McEuen, P. L. Fabrication of Metallic Electrodes with Nanometer Separation by Electromigration. *Appl. Phys. Lett.* **1999**, *75* (2), 301–303

(2) Cui, L.; Zhu, Y.; Nordlander, P.; Ventra, M. Di; Natelson, D. Thousand-Fold Increase in Plasmonic Light Emission via Combined Electronic and Optical Excitations. *Nano Lett.* **2021**, *21* (6), 2658–2665

(3) Cui, L.; Zhu, Y.; Abbasi, M.; Ahmadivand, A.; Gerislioglu, B.; Nordlander, P.; Natelson, D. Electrically Driven Hot-Carrier Generation and Above-Threshold Light Emission in Plasmonic Tunnel Junctions. *Nano Lett.* **2020**, *20* (8), 6067–75

(4) Bolotin, K. I.; Kuemmeth, F.; Pasupathy, A. N.; Ralph, D. C. Metal-Nanoparticle Single-Electron Transistors Fabricated Using Electromigration. *Appl. Phys. Lett.* **2004**, *84* (16), 3154

(5) Buret, M.; Smetanin, I. V.; Uskov, A. V.; Colas Des Francs, G.; Bouhelier, A. Effect of Quantized Conductivity on the Anomalous Photon Emission Radiated from Atomic-Size Point Contacts. *Nanophotonics* **2020**, *9* (2), 413–425

(6) Zhang, C.; Hugonin, J.-P.; Coutrot, A.-L.; Sauvan, C.; Marquier, F.; Greffet, J.-J. Antenna Surface Plasmon Emission by Inelastic Tunneling. *Nat. Commun.* **2019**, *10* (1), 4949

(7) Kalathingal, V.; Dawson, P.; Mitra, J. Scanning Tunnelling Microscope Light Emission: Finite Temperature Current Noise and over Cut-off Emission. *Sci. Rep.* **2017**, *7* (1), 3530

(8) Peters, P.-J.; Xu, F.; Kaasbjerg, K.; Rastelli, G.; Belzig, W.; Berndt, R. Quantum Coherent Multielectron Processes in an Atomic Scale Contact. *Phys. Rev. Lett.* **2017**, *119* (6), 066803

(9) Kaasbjerg, K.; Nitzan, A. Theory of Light Emission from Quantum Noise in Plasmonic Contacts: Above-Threshold Emission from Higher-Order Electron-Plasmon Scattering. *Phys. Rev. Lett.* **2015**, *114* (12), 126803

(10) Février, P.; Gabelli, J. Tunneling Time Probed by Quantum Shot Noise. *Nat. Commun.* **2018**, *9* (1), 4940

(11) Fung, E. D.; Venkataraman, L. Too Cool for Blackbody Radiation: Overbias Photon Emission in Ambient STM Due to Multielectron Processes. *Nano Lett.* **2020**, *20* (12), 8912–8918

(12) Xu, F.; Holmqvist, C.; Belzig, W. Overbias Light Emission Due to Higher-Order Quantum Noise in a Tunnel Junction. *Phys. Rev. Lett.* **2014**, *113* (6), 066801

(13) Xu, F.; Holmqvist, C.; Rastelli, G.; Belzig, W. Dynamical Coulomb Blockade Theory of Plasmon-Mediated Light Emission from a Tunnel Junction. *Phys. Rev. B* **2016**, *94*, 245111

(14) Schneider, N. L.; Johansson, P.; Berndt, R. Hot Electron Cascades in the Scanning Tunneling Microscope. *Phys. Rev. B* **2013**, *87* (4), 45409

(15) Herzog, J. B.; Knight, M. W.; Li, Y.; Evans, K. M.; Halas, N. J.; Natelson, D. Dark Plasmons in Hot Spot Generation and Polarization in Interelectrode Nanoscale Junctions. *Nano Lett.* **2013**, *13* (3), 1359–1364

(16) Rakić, A. D. Algorithm for the Determination of Intrinsic Optical Constants of Metal Films: Application to Aluminum. *Appl. Opt.* **1995**, *34* (22), 4755

(17) Buret, M.; Uskov, A. V.; Dellinger, J.; Cazier, N.; Mennemanteuil, M.-M.; Berthelot, J.; Smetanin, I. V.; Protsenko, I. E.; Colas-des-Francs, G.; Bouhelier, A. Spontaneous Hot-Electron Light Emission from Electron-Fed Optical Antennas. *Nano Lett.* **2015**, *15* (9), 5811–5818

(18) Simmons, J. G. Generalized Formula for the Electric Tunnel Effect between Similar Electrodes Separated by a Thin Insulating Film. *J. Appl. Phys.* **1963**, *34* (6), 1793–1803

(19) Downes, A.; Dumas, P.; Welland, M. E. Measurement of High Electron Temperatures in Single Atom Metal Point Contacts by Light Emission. *Appl. Phys. Lett.* **2002**, *81* (7), 1252–1254

(20) Pechou, R.; Coratger, R.; Ajustron, F.; Beauvillain, J. Cutoff Anomalies in Light Emitted from the Tunneling Junction of a Scanning Tunneling Microscope in Air. *Appl. Phys. Lett.* **1998**, *72* (6), 671–673

(21) Malinowski, T.; Klein, H. R.; Iazykov, M.; Dumas, P. Infrared Light Emission from Nano Hot Electron Gas Created in Atomic Point Contacts. *EPL* **2016**, *114* (5), 57002

(22) Shalem, G.; Erez-Cohen, O.; Mahalu, D.; Bar-Joseph, I. Light Emission in Metal–Semiconductor Tunnel Junctions: Direct Evidence for Electron Heating by Plasmon Decay. *Nano Lett.* **2021**, *21* (3), 1282–1287

(23) Ott, C.; Götzinger, S.; Weber, H. B. Thermal Origin of Light Emission in Nonresonant and Resonant Nanojunctions. *Phys. Rev. Res.* **2020**, *2* (4), 042019

(24) Leon, C. C.; Rosławska, A.; Grewal, A.; Gunnarsson, O.; Kuhnke, K.; Kern, K. Photon Superbunching from a Generic Tunnel Junction. *Sci. Adv.* **2019**, *5* (5), eaav4986

(25) Martín-Jiménez, A.; Lauwaet, K.; Jover, Ó.; Granados, D.; Arnau, A.; Silkin, V. M.; Miranda, R.; Otero, R. Electronic Temperature and Two-Electron Processes in Overbias Plasmonic Emission from Tunnel Junctions. *Nano Lett.* **2021**, *21* (16), 7086–7092